\documentclass{article}
\usepackage{frascatiphys,here,graphicx,subfigure}
\begin{document}
\title{Exotica possibility of new observations by BES}
\author{
Ailin Zhang \\
{\em Department of Physics, Shanghai University,
Shanghai, 200444, China} } \maketitle \baselineskip=11.6pt
\begin{abstract}
The employment of interpolating currents of existed studies of
four-quark state and glueball with QCD sum rule approach is
analyzed. In terms of suitable currents, the masses of the lowest
lying scalar and pseudo-scalar glueball were determined. The masses
of some tetraquark states and their first orbital excitations were
obtained through a combination of the sum rule with the constituent
quark model. Exotica possibility of the new observations by BES is
discussed.
\end{abstract}
\baselineskip=14pt
\section{QCD sum rules and exotica}
QCD is believed the right theory describing strong interactions,
quark model is proved successful in describing normal hadron.
However, the low energy behavior of QCD and the mechanism of quark
confinement of hadron are not clear. The study of hadron properties
with QCD is a great challenge. In history, many models based on QCD
were developed to study hadron. QCD (SVZ) sum rule~\cite{svz} is
such an effective nonperturbative method of relating fundamental
parameters of QCD Lagrangian and vacuum to parameters of hadrons.

In sum rules method, to detect the properties of hadrons, some
correlators are constructed from suitable interpolating currents
(local operators). In one hand, the correlator is expanded in
perturbative coefficients and condensates. In the other hand, the
imaginary part of the correlator (spectral density) is expressed
with the parameters of resonances. Through a dispersion relation,
the parameters of QCD and vacuum are connected with the parameters
of hadrons. To get reasonable conclusions on the properties of
hadrons, it is very important to employ suitable currents.

In normal hadron case, the structure of meson and baryon is not so
complex and sum rule works well. Exotic hadrons such as glueball,
hybrid and multi-quark state have complex intrinsic structure, there
are often different ways to employ interpolating currents.
Furthermore, no exotic hadron has been confirmed such that the
properties of exotic hadrons are not clear. Whether QCD sum rule
works or not has not been proved. It should be careful to employ
reasonable currents and to draw corresponding conclusions on exotic
hadrons.
\section{$0^{++}$ and $0^{-+}$ glueballs}
The existence of glueball was firstly mentioned by Fritzsch and
Gell-Mann~\cite{fritzsch}. Glueball was studied in many models. In
sum rules approach, the interpolating currents consist of gluons
fields.

For $0^{++}$ glueball, the current was firstly employed by Novikov
et al\cite{nsvz}.,
\begin{eqnarray}
j_s=\alpha_s G^a_{\mu\nu}G^a_{\mu\nu},
\end{eqnarray}
where $m_\sigma=700$ MeV was taken as that of the $\sigma$ without
computation.

Subsequently, the $0^{++}$ scalar glueball was studied with the same
current in many literatures. In Narison's~\cite{narison} work,
$m=1.5\pm 0.2$ GeV was predicted. In the work of
Huang's~\cite{huang}, $m=1.7\pm 0.2$ GeV was predicted with a
reasonable moment. In the work of Harnett's~\cite{harnett}, two
glueballs were predicted with the heavier: $m=1.4$ GeV and the
lighter: $m\approx 1.0-1.25$ GeV, where the contribution of
instanton was taken into account. In a most recent work by
Forkel~\cite{forkel}, with a comprehensive inclusion of the
contribution of operator product expansion, $m=1.25\pm 0.2$ GeV was
predicted.

For $0^{-+}$ glueball, the interpolating current was also firstly
employed by Novikov et al\cite{nsvz2}.,
\begin{eqnarray}
j_{ps}=\alpha_s G^a_{\mu\nu}\tilde G^a_{\mu\nu}.
\end{eqnarray}
In their computation, $m=2-2.5$ GeV. This interpolating current was
also employed to study pseudoscalar glueball in many other
literatures. In Narison's~\cite{narison} work, $m=2.05\pm 0.19$ GeV
was predicted. With the higher-loop perturbative contributions and
instantons taken into account, $m=2.65\pm 0.33$ GeV was predicted in
the work of Zhang's~\cite{zhang}. In Forkel's work, the instanton
and the topological charge screening effect were taken into account,
and $m=2.2\pm 0.2$ GeV. It is widely believed that the mass of the
$0^{-+}$ glueball is larger than that of the $0^{++}$ glueball.

In the constituent parton model, there are glueball with two gluons
and glueball with three gluons. In sum rule method, in addition to
the interpolating currents consisting of two gluons field,
interpolating currents consisting of three gluons field have been
employed. There is a large mass difference between the $0^{-+}$ and
the $0^{++}$ glueball prediction. The difference may results from
the rough calculations or the special features of the $0^{-+}$
glueball and the $0^{++}$ glueball. For the difficulty in the
calculation of the OPE, present results on glueball masses are not
definite and may be improved largely with more accurate computation.

Current consisting of three gluons field was firstly employed by
Latorre et al.~\cite{latorre} to compute the mass of the $0^{++}$
scalar three gluons glueball
\begin{eqnarray} j_{s3g}=g^3f_{abc}
G^a_{\mu\nu}G^b_{\nu\rho}G^c_{\rho\mu},
\end{eqnarray}
with $m_{s3g}=3.1$ GeV.

Current consisting of three gluons field was recently employed by
Hao et al.~\cite{hao} to compute the mass of the $0^{-+}$
pseudoscalar three gluons glueball
\begin{eqnarray}
j_{ps3g}=g^3f_{abc} \tilde G^a_{\mu\nu}\tilde G^b_{\nu\rho}\tilde
G^c_{\rho\mu},
\end{eqnarray}
with $m_{ps3g}=1.9-2.7$ GeV.

As well known, two gluons glueball may mix with three gluons
glueball, and two gluons currents may mix with three gluons
currents. Furthermore, this two kinds of currents couple to both
kinds of glueballs. How to deal with these mixing effects is a great
challenge in sum rule approach. Final conclusions on glueball are
expected to depend heavily on these mixing effects.

\section{$0^{++}$ and $1^{--}$ tetraquark states}
Four-quark state has been studied in MIT bag model~\cite{bag}, color
junction model~\cite{junction}, potential model~\cite{potential},
effective Lagrangian method~\cite{lag}, relativistic quark
model~\cite{rquark}, QCD sum
rules~\cite{sum1,sum2,sum3,sum4,sum5,sum6} and many other
methods~\cite{other}. More references could be found in Refs.
~\cite{review} and therein.

Four-quark state consists of four quarks and anti-quarks. Their
intrinsic quarks/anti-quarks may make different clusters
(correlations) such as color, flavor, spin, {\it
etc}~\cite{bag,sum6,review}. According to the spacial extension of
clusters, there are two different types of four-quark states:
$(qq)(\bar q\bar q)$ and $(q\bar q)(q\bar q)$. $(qq)(\bar q\bar q)$
is often called tetraquark state or baryonium, which consists of
diquark $qq$ and anti-diquark $\bar q\bar q$. $(q\bar q)(q\bar q)$
includes the molecule state.

$(qq)(\bar q\bar q)$ and $(q\bar q)(q\bar q)$ may mix with each
other, and they may mix with normal meson $q\bar q$ (the ones mixed
with $q\bar q$ are usually called crypto-exotic four-quark states).
Therefore, the meson observed by experiment is a mixed one
\begin{eqnarray}
|meson>=|q\bar q>+|(qq)(\bar q\bar q)>+|(q\bar q)(q\bar q)>+\cdots.
\end{eqnarray}

Quark dynamics in four-quark state is still not clear, so intrinsic
color, flavor configurations in four-quark state could not be
distinguished except that some special observable is established.
Unfortunately, no such an observable has been definitely set up.

To study four-quark state with sum rule, two kinds of interpolating
currents, for example,$(q\bar q)(q\bar q)$~\cite{sum1}, $(q\bar
q)^2$, $(qq)(\bar q\bar q)$~\cite{sum2}, $(cq)(\bar q\bar
q)$~\cite{sum3}, $(cu)(\bar s\bar u)$~\cite{sum4}, $(ud)(\bar s\bar
s)$~\cite{sum5}, have been employed. All the calculations are in
leading order.

Many conclusions on four-quark states have been drawn based on this
two kinds currents. However, in view of the sum rule approach, there
is no definite difference between this two kinds currents. The
reason is that $(qq)(\bar q\bar q)$ and $(q\bar q)(q\bar q)$ can be
turned into each other after Fierz transformation~\cite{sum2,sum6},
and they will mix with each other under renormalization. Therefore,
it is useful to remember that conclusions on the structure of
four-quark state in constituent quark picture can not be drawn
directly from the structure in current (operator) picture.
Similarly, diquark concept is not meaningful in current
picture~\cite{sum6}. In principle, there is no direct way to turn
the current (operator) picture into the constituent quark picture.

To get a reasonable result on four-quark state, suitable mixed
interpolating currents and mixture of hadrons should be taken into
account, which is also a great challenge in sum rule method.

Following the diquark picture applied to weak hadron decays with sum
rules~\cite{jamin}, the diquark current with flavor $(sq)$
\begin{eqnarray}
j_i(x)=\epsilon_{ijk}s^T_j(x)COq_k(x)
\end{eqnarray}
was employed and an updated analysis was performed in a recent
attempt~\cite{sum6}. The most "suitable" masses of diquark $m_{qq}$
and $m_{sq}$ were obtained: $m_{qq}\sim 400$ MeV and $m_{sq}\sim
460$ MeV with $s_0=1.2$ GeV$^2$. The mass scale of diquark is the
same as that of the constituent quark. The results obtained here are
consistent with the fit of Maiani's~\cite{other}.

Once the masses determined by sum rule are taken as the constituent
diquark masses, masses of the $L=0$ and the $L=1$ excited tetraquark
state are obtained as the method of Maiani's~\cite{other}
\begin{eqnarray*}
M\approx 2m_{[qq]}-3(\kappa_{qq})_{\bar 3},~~M\approx
2m_{[qq]}-3(\kappa_{qq})_{\bar 3}+B^\prime_q{L(L+1)\over 2}.
\end{eqnarray*}
The obtained masses of some four-quark states are listed in
tab.\ref{tab1}. Tetraquark states consisting of bad diquark have the
same mass scale~\cite{other}. It is easy to find the explicit flavor
dependence of masses.
\begin{table}[t]
\centering \caption{ \it Masses of some tetraquark states. } \vskip
0.1 in
\begin{tabular}{|l|c|c|} \hline
          &  $0^{++}$ & $1^{-+}$ \\
\hline \hline $[qq][\bar q\bar q]$  & $\sim490$~~MeV &
$\sim 490+B^\prime_q$~~MeV    \\
$[sq][\bar q\bar q]$   & $\sim 610$~~MeV  &
$\sim 610+B^\prime_q$~~MeV     \\
$[sq][\bar s\bar q]$   & $\sim 730$~~MeV  &
$\sim 730+B^\prime_s$~~MeV      \\
\hline
\end{tabular}
\label{tab1}
\end{table}
\section{Exotica possibility of the new observations by BES}
Some new observations were reported by BES through its 58 million
events sample of $J/\Psi$ decays.

$p\bar p$ enhancement was observed by BES~\cite{bes1} in the
radiative decay $J/\Psi\to\gamma p\bar p$ with
$M=1859^{+3}_{-10}(stat)^{+5}_{-25}(sys)$ (below $2m_p$) and
$\Gamma<30$ MeV if interpreted as a single $0^{-+}$ resonance. It
was observed also by Belle~\cite{belle} and BaBar~\cite{babar}
collaborations in other channels. Its quantum number assignment
$J^{PC}$ is consistent with either $0^{-+}$ or $0^{++}$. This
enhancement was once interpreted as final state interaction effect,
baryonium or threshold cusp.

$X(1835)$ was observed by BES~\cite{bes2} in the decay
$J/\Psi\to\gamma\pi^+\pi^-\eta^\prime$ with $M=1833.7\pm
6.1(stat)\pm 2.7(syst)$ MeV and $\Gamma=67.7\pm 20.3\pm 7.7$ MeV. It
is consistent with expectations for the state that produces the
strong $p\bar p$ mass threshold enhancement. It was once interpreted
as $0^{-+}$ glueball or baryonium.

$X(1812)$ was observed by BES~\cite{bes3} in the doubly
OZI-suppressed decay $J/\Psi\to\gamma\omega\phi$ with
$M=1812^{+19}_{-26}(stat)\pm 18(syst)$ MeV, $\Gamma=105\pm 20\pm 28$
MeV. It favors $J^P=0^+$. It was interpreted as rescatterings
effect, four-quark state, glueball or hybrid.

$X(1576)$ was observed by BES~\cite{bes4} in the decay $J/\Psi\to
K^+K^-\pi^0$ with pole position
$1576^{+49}_{-55}(stat)^{+98}_{-91}(syst)$
MeV-$i(409^{+11}_{-12}(stat)^{+32}_{-67})$ MeV. This broad peak is
believed to have $J^{PC}=1^{--}$. It was interpreted as final state
interaction effect or tetraquark state.

Exotica was often invoked to explain the special features of newly
observed states. Based on previous analysis, the glueball and
tetraquark possibility of these observations is examined.

In QCD sum rule approach, the two glueball candidates with lower
mass are $0^{++}$ and $0^{-+}$ glueball. Therefore, $X(1835)$ and
$X(1812)$ may be a $0^{++}$ glueball, but they are unlike to be the
pure $0^{-+}$ glueball.

$0^{++}$ and $0^{-+}$ tetraquark states have the same mass scale,
and they have lower masses compared with the new observations by
BES. It is hard to explain these observations as $0^{++}$ or
$0^{-+}$ tetraquark state. If $X(1576)$ is confirmed to have
$1^{--}$, it may be the first orbital excited $1^{--}$ tetraquark
state (orbital excitation of $a_0(980)$ or $f_0(980)$) with a very
large excited energy $\sim 596$ MeV.

\section{Conclusions and discussions}
Different interpolating currents have been employed to study exotic
states, but the structure of interpolating currents has no direct
correspondence to the constituent structure of hadrons. The study of
exotica with sum rule requires more exploration.

Masses of $0^{++}$, $0^{-+}$ glueballs and some tetraquark states
were determined. Through these studies, the new observations by BES
are unlike to be the pure $0^{-+}$ pseuso-scalar glueball, they are
unlikely to be the light tetraquark states either except that
$X(1576)$ may be an exotic first orbital excited $(sq)(\bar s\bar
q)$ tetraquark state.

\section{Acknowledgements}
This work is supported in part by NSFC under the grant: 10775093.
The author thanks all the collaborators and is grateful to the
organizing committee for partial financial support.

\end{document}